\documentclass[preprint,aps,prd,showpacs,nofootinbib]{revtex4}
\usepackage{mathrsfs}
\usepackage{amsmath}
\usepackage{graphicx}
\usepackage{subfigure}

\newcommand{\bea}{\begin{eqnarray}}
\newcommand{\eea}{\end{eqnarray}}
\newcommand{\beq}{\begin{equation}}
\newcommand{\eeq}{\end{equation}}

\def\/{\over}

\def\gsim{ \lower .75ex \hbox{$\sim$} \llap{\raise .27ex \hbox{$>$}} }
\def\lsim{ \lower .75ex \hbox{$\sim$} \llap{\raise .27ex \hbox{$<$}} }

\begin{document}

\title{A new extended quintessence}

\author{Peng Wang, Puxun Wu and  Hongwei Yu }
\address
{Center of Nonlinear Science and Department of Physics, Ningbo
University,  Ningbo, Zhejiang, 315211 China }

\begin{abstract}
The extended quintessence is obtained by coupling a normal scalar field to the Ricci scalar  defined in the metric formalism.
In this paper, we propose a new extended quintessence dark energy by introducing a non-minimal coupling between the quintessence  and gravity, but with the Ricci scalar given from the Palatini formalism rather than the metric one. We find that the equation of state of the new extended quintessence can cross the phantom divide line, and moreover, it oscillates around the $-1$ line.
 We also show that the universe driven by the new extended quintessence will enter a dark energy dominated de Sitter phase in the future.

\end{abstract}

\maketitle

\section{Introduction}\label{sec1}
Various observational data, including the Type Ia supernovae~\cite{Perlmutter1999, Riess1998},  the cosmic microwave background radiation~\cite{Spergel} and the large scale structure~\cite{Eisenstein,Tegmark}, and so on,  have confirmed that our universe is undergoing an  accelerating expansion. To explain this observed phenomenon, usually, an exotic energy component, named dark energy, is assumed to exist in our universe. The simplest candidate of dark energy is the cosmological constant. Although it can explain the observation fairly well, it suffers from both the fine-tuning  problem and the coincidence problem~\cite{CC}. Moreover, recent observation seems to favor a dynamical dark energy~\cite{Zhao}. Thus, some minimally  coupled scalar field dark energy models, such as quintessence, phantom and quintom, are proposed to explain the observed accelerating expansion.  The quintessence~\cite{Quintessence} is a normal scalar field and its equation of state $w$ is larger than $-1$. The phantom~\cite{Phantom} has a negative kinetic term such that its $w$ is less than $-1$. Combining the quintessence and phantom in a unified model, the quintom~\cite{Quintom}  is obtained. The advantage of the quintom is that it  can realize the crossing of the $-1$ line for  the equation of state.

 Since non-minimal couplings are generated by quantum corrections  and they are essential for the renormalizability of scalar field theory in curved space, it is natural to consider  a  coupling between the scalar field  and gravity, $\omega \phi^2 R$, where $R$ is the Ricci scalar. This coupling has been widely studied  for both inflation and dark energy, and
it has some novel properties, such as inflation can be easily obtained \cite{Abbott}, and  the effective equation of state of the non-minimal coupling  quintessence can cross the phantom divide line~\cite{Sahni}.  When dark energy is concerned, the model with a coupling of the normal scalar field and gravity is   known as the extended quintessence (EQ).

It is well known that the general relativity field equation can be obtained not only in the metric formalism but also in the Palatini formalism~\cite{Misner}. In the former case, only the metric is  variable, while in the latter case, both the metric and the connection are independent  variables.  When the Einstein-Hilbert action is generalized to be an arbitrary function $f$ of $R$, the Palatini formalism leads to second order differential equations instead of the fourth order ones that one gets with the metric variation~(see \cite{Sotiriou,Felice} for recent reviews). Thus, the Palatini modified gravity  theory seems to have an advantage since it can pass the solar system tests automatically.

Recently, Geng et al found that, when a non-minimal coupling between the quintessence and teleparallel gravity  is switched on,    the resulting theory has a richer structure than the extended quintessence  and named it teleparallel dark energy~\cite{Geng}, although,  in the minimal coupling case, the quintessence scalar field cosmological model in teleparallel gravity  is identical to that in general relativity. In this paper, we plan to explore what happens when a non-minimal coupling is introduced between the quintessence and the Ricci scalar $\hat R$  in the  Palatini formalism (Hereafter, an {\it overhat}  denotes that the quantity is defined with the independent Palatini connection) and we call this coupled scalar field  ``new extended quintessence".

The paper is organized as follows. In Sec.~II,  we  derive the field equation in the case of a non-minimal coupling between the quintessence and $\hat R$.
In Sec.~III, we discuss  the evolutionary curve of the effective equation of state and in Sec.~IV the dynamical behavior of the system. In Sec.~V we  conclude. Throughout the paper, unless otherwise specified, we work in units in which $c = 8\pi G = 1$. Greek and Latin indices run from $0$ to $3$ and $1$ to $3$, respectively.

\section{field equation in palatini gravity with a non-minimal coupling}

 We consider an action that contains a non-minimal coupling between the quintessence scalar field and the Palatini Ricci scalar:
\bea\label{action}
S=\int dx^4 \sqrt{-g} \bigg[{\frac{1}{2}\hat{R}(g_{\mu\nu}, \hat{\Gamma}^\alpha_{\beta\gamma})}+\frac{1}{2}\omega\hat{R}\phi^2+\mathcal{L}_\phi+\mathcal{L}_m\bigg]\;, \eea
where $\omega$ is the coupling constant, and the metric $g_{\mu\nu}$ and the connection $\hat{\Gamma}^\alpha_{\beta\gamma}$  are two independent  variables. $\mathcal{L}_m$ is the ordinary matter Lagrangian and $\mathcal{L}_\phi$ is the Lagrangian of the quintessence  field, which is taken to be\bea \mathcal{L}_\phi= -\frac{1}{2}\partial_\mu\phi\partial^\mu\phi-V(\phi) \eea
 The Ricci scalar $\hat{R}$ is obtained by contracting the Ricci tensor $\hat{R}_{\mu\nu}$, which is defined in terms of the independent connection $\hat{\Gamma}^\alpha_{\beta\gamma}$
 \bea \hat{R}_{\mu\nu}=\hat{\Gamma}^\alpha_{\mu\nu,\alpha}-\hat{\Gamma}^\alpha_{\mu\alpha,\nu}+\hat{\Gamma}^\alpha_{\alpha\lambda}\hat{\Gamma}^\lambda_{\mu\nu}-\hat{\Gamma}^\alpha_{\mu\lambda}\hat{\Gamma}^\lambda_{\alpha\nu}\;.\eea
Varying the action (\ref{action}) with respect to the metric $g_{\mu\nu}$ and the connection $\hat\Gamma^\lambda_{\mu\nu}$, respectively, one obtains
\bea \label{Eqf} F \hat R_{\mu\nu}-{1\/2}f(\hat R, \phi)g_{\mu\nu}= T_{\mu\nu} ,\eea
and
\bea\label{Eqc} \hat\nabla_\lambda(\sqrt{-g}F g^{\mu\nu})=0. \eea
Here  $F\equiv 1+\omega\phi^2$, $f\equiv F \hat R$ and $T_{\mu\nu}$ is the energy-momentum tensor including usual matter and the scalar field.
Contracting Eq.~(\ref{Eqf}), one can get: \bea F \hat R-2f=-f= T\;,\eea
where $T=-\rho_m-\rho_\phi+3(p_m+p_\phi)$ with $\rho_m$ and $\rho_\phi$ being the energy densities of matter and the scalar field, respectively, and $p_m$ and $p_{\phi}$ being the corresponding pressures. Using $p_m=0$,  $\rho_\phi=\frac{1}{2}\dot{\phi}^2+V$ and $p_\phi=\frac{1}{2}\dot{\phi}^2-V$, one has $ T =-\rho_m+\dot\phi^2-4V$.

From Eq.~(\ref{Eqc}), one can define a new metric $h_{\mu\nu}$ conformally connected to $g_{\mu\nu}$ by $h_{\mu\nu}=Fg_{\mu\nu}$. Then Eq.~(\ref{Eqc}) becomes
\bea \hat\nabla_\lambda(\sqrt{-h}h^{\mu\nu})=0\;,\eea which means that $\hat\Gamma^\lambda_{\mu\nu}$ can be expressed as the Levi-Civita connection with respect to the new metric  $h_{\mu\nu}$
 \bea \hat\Gamma^\lambda_{\mu\nu}={1\/2}h^{\lambda\rho}(h_{\mu\rho,\nu}+h_{\nu\rho,\mu}-h_{\mu\nu,\rho})\;. \eea
 Using the above equation, we can derive the expressions  of the Palatini Ricci tensor and  the Ricci scalar:
 \bea\label{Edm1} \hat R_{\mu\nu}(\Gamma)&=&R_{\mu\nu}(g)+{3\/2} {\nabla_\mu F\nabla_\nu F\/F^2}-{\nabla_{\mu}\nabla_\nu F\/F}-g_{\mu\nu}{\nabla_\sigma\nabla^\sigma F\/{2F}}\;,\eea
  \bea\label{Edm2} \hat R(\Gamma)&=&R(g)+{3\/2}{\nabla_\sigma F\nabla^\sigma F\/F^2}-3{\nabla_\sigma\nabla^\sigma F\/F}\;,\eea
 where $R_{\mu\nu}$ and $R$ are the Ricci tensor and the Ricci scalar defined  in the metric formalism, respectively, and $\nabla_\sigma$ is a covariant derivative associated with the usual Levi-Civita connection.

Now, we consider a flat Friedmann-Lemaitre-Robertson-Walker (FLRW) universe. The metric has the form:
\bea ds^2=-dt^2+a^2dx_idx_j\delta^{ij}  \eea
where $a(t)$ is the cosmic scale factor and $t$ is the cosmic time. Using Eqs.~(\ref{Eqf}, \ref{Edm1}, \ref{Edm2}), we obtain the modified Friedmann equation and Raychaudhuri equation:
\bea H^2={1\/3}\rho_m+\frac{1}{3}\bigg({1\/2}{\dot\phi}^2+V\bigg)-H\dot F-\frac{1}{4}{\dot{F}^2\/{F}}-\omega H^2\phi^2\;, \eea
\bea  -3H^2-2\dot{H}={1\/2}\dot\phi^2-V+2H\dot{F}+\ddot{F}-{3\/4}{\dot{F}^2\/F}+\omega \phi^2(3H^2+2\dot{H})\;. \eea
Combining the above two equations leads to \bea{\dot{H}\/H^2}=-{1\/2FH^2}(\rho_m+\dot\phi^2)+{3\/4}\bigg({\dot{F}\/H F}\bigg)^2+{1\/2}{\dot{F}\/H F}-\frac{1}{2} {\ddot{F}\/FH^2}\;.  \eea
Writing the modified Friedmann equation and Raychaudhuri equation in the standard form, one can define an effective energy density and pressure for  the non-minimally coupled quintessence
 \bea\label{rhoe} \rho_{\phi,eff}=\frac{1}{2}\dot{\phi}^2+V-3H\dot{F}-\frac{3}{4}\frac{\dot{F}^2}{F}-3\omega H^2\phi^2\;, \eea
 \bea\label{pe} p_{\phi,eff}=\frac{1}{2}\dot{\phi}^2-V+2H\dot{F}+\ddot{F}-\frac{3}{4}\frac{\dot{F}^2}{F}+
 \omega\phi^2(3H^2+2\dot{H}) \;,\eea
which  differ from the non-minimal coupling case in the metric formalism. When $\omega=0$, the above two equations reduce to the usual ones. Using Eq.~(\ref{rhoe}), one can define the dimensionless density parameters  for dark energy and matter:
\bea \Omega_{DE}=\frac{{\rho}_{\phi,eff}}{3H^2}\;,\;\;\; \Omega_{m}=1-  \Omega_{DE}=\frac{\rho_m}{3H^2}\;.\eea
Eqs.~(\ref{rhoe}) and (\ref{pe}) show that  ${\rho}_{\phi,eff}$ and $p_{\phi,eff}$  satisfy the normal  energy conservation equation
 \bea\label{Eqco} \dot{\rho}_{\phi,eff}+3H(\rho_{\phi,eff}+p_{\phi,eff})=0\;. \eea
From Eq.~(\ref{Eqco}), we find the field equation of the quintessence with a non-minimal
coupling
\bea \ddot{\phi}+3H\dot{\phi}+V_{,\phi}-\omega\phi \hat R=0\;. \eea
This equation can also be obtained by varying action (\ref{action}) with respect to the scalar field.

\section{evolutionary curve of the effective equation of state}
\begin{figure}[htbp]
\centering
\includegraphics[scale=0.8]{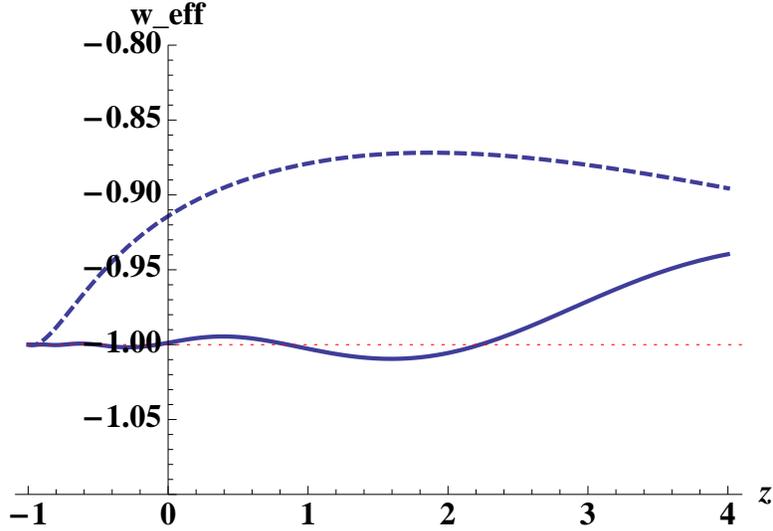}
\caption{\label{Fig1}Evolution of the dark energy equation of state parameter $w_{eff}$ as a function of the redshift $z$.  An exponential scalar field potential $V\propto e^{\lambda \phi}$ is considered. The dashed curve presents quintessence-like behavior and corresponds to $\omega=-0.16$ and $\lambda=1$, and the solid curve presents an oscillation behavior with the phantom divide line crossing and corresponds to $\omega=-2$ and $\lambda=1$. }
\end{figure}

Using the effective energy density  and pressure, one can define an effective equation of state for the new extended quintessence dark energy:
\bea w_{eff}=\frac{ \rho_{\phi,eff}}{ p_{\phi,eff}}\;.\eea
In Fig.~(\ref{Fig1}), we plot the evolutionary curves of $w_{eff}$ in an exponential scalar field potential case with different values of  model parameters. The dashed curve shows that the new extended quintessence dark energy behaves like quintessence when the coupling is weak (small $|\omega|$), while, in the case of a solid curve, it exhibits an oscillatory behavior with  crossings of the phantom divide line when the coupling is strong ($|\omega|$ of the order of unity).  The oscillation is a distinctive feature of the new extended quintessence dark energy in contrast to the extended quintessence~\cite{Sahni} and teleparallel dark energy~\cite{Geng} where the scalar field is coupled to the metric Ricci scalar and the torsion scalar, respectively. Notice, however, that   an oscillating equation of state can also be obtained in a viable $f(R)$ theory ~\cite{Bamba}. It is interesting to  note that $w_{eff}\rightarrow-1$ in the future,, which means that  the universe will enter a dark energy dominated de Sitter phase finally. 

\section{dynamic behavior }
In this Section, we plan to discuss the dynamical behavior of  the system discussed in the second Section.   For convenience, we rewrite the Friedmann equation  as :
\bea 1=\frac{1}{3}{\rho_m\/H^2}+\frac{1}{6}\frac{\dot\phi^2}{H^2}+\frac{1}{3}{V\/H^2}-{\dot{F}\/H}- \frac{1}{4}{\dot{F}^2\/H^2F}
-\omega\phi^2\;. \eea
Following e.g. \cite{Wei}, we introduce the following dimensionless variables:
\bea x_1=\frac{\dot\phi}{\sqrt{6}H},\;\;\; x_2={\sqrt{V}\/\sqrt{3}H}, \;\;\;x_3={\sqrt{\rho_m}\/\sqrt{3}H},\;\;\;x_4=\phi\;. \eea
Apparently, only three of them are independent since the Friedmann equation gives a constraint
\bea x_1^2+x_2^2+x_3^2-2\sqrt{6}\omega x_1 x_4-\omega x_4^2-6\omega\frac{x_1^2x_4^2}{1+\omega x_4^2}=1\;. \eea

For simplicity, an   exponential potential: ${V_{,\phi}\/ V}=\lambda$ is considered in our discussion. Using the energy conservation equations of matter and the scalar field,  the modified  Friedmann equation, and Raychaudhuri equation, we find that the dynamical system can be expressed as
\bea {d {x_1}\/dN}&=&{c\/\sqrt 6}- s{x_1}\;,\\
{d {x_2}\/dN}&=&{ \sqrt 6\/2}\lambda {x_2}{x_1}-s{x_2}\;,\\
{d x_3\/dN}&=&-\frac{3}{2}{x_3}-s {x_3}\;,\\
{d x_4\/dN}&=&\sqrt 6{x_1}\;,\eea
where  $ N\equiv \ln{a}$, $c$ and $s$ are defined as
\bea c\equiv-3 \sqrt{6} {x_1}-3 \lambda x_2^2 +\frac{  \omega{x_4}}{1+\omega x_4^2  }  (-6 x_1^2+12 x_2^2+3 {x_3^2})\;,
\eea
\bea s\equiv{\dot{H}\/H^2}=-\frac{3}{2}{{x_3^2}\/F}-3{x_1^2\/F}-6\omega  {x_1^2\/F}-\omega c{ {x_4}  \/F}+18\omega^2 \bigg(\frac{ {x{_1}} {x{_4}}  }{1+\omega x_4^2}\bigg)^2+  \sqrt{6} \omega \frac{ {x{_1}} {x{_4}}  }{1+\omega x_4^2}\;,   \eea with $ F=1+\omega  x_4^2 $.

The critical points  ($\bar{x}_1,\bar{x}_2, \bar{x}_3, \bar{x}_4$) of  the autonomous system (Eq.(24)-Eq.(27)) can be obtained through the conditions \bea \frac{dx_1}{dN} = \frac{dx_2}{dN} =\frac{dx_3}{dN}=\frac{dx_4}{dN}=0.\eea Then we get  three critical points  which are given along with  their corresponding existence conditions in Tab.~(\ref{Tab1}). We find that, for Points (C.P.a) and (C.P.b), $\Omega_{DE}=1$, $\Omega_m=0$ and $w_{eff}=-1$, which means that both of them are dark energy dominated de Sitter solutions. Apparently, Point (C.P.c) is a matter dominated solution.

\begin{table}[!h]\centering
\begin{tabular}{|c|c|c|c|c|c|c|c|}
\hline
Label &~Critical Point ($\bar{x}_1,\bar{x}_2, \bar{x}_3, \bar{x}_4$)~&~Existence~& $w_{eff}$\\
\hline
C.P.a & $ 0,\;2 \sqrt{\left(2 \omega+\sqrt{-\lambda ^2\omega+4 \omega^2  }\right){\lambda^{-2}} },\;0,\;\frac{2 \omega +\sqrt{-\lambda ^2 \omega +4 \omega ^2}}{\lambda  \omega } $& $ \lambda^2\leq 4\omega$ or $\omega<0$ & $-1$\\
\hline
C.P.b  &$0,\; 2\sqrt{\left(2 \omega -\sqrt{-\lambda ^2\omega+4 \omega^2}\right)\lambda^{-2} },\; 0,\; \frac{2 \omega -\sqrt{-\lambda ^2 \omega +4 \omega ^2}}{\lambda  \omega }$  & $ \lambda^2\leq 4\omega$  & $-1$\\
\hline
C.P.c    &$0,\;0,\;1,\;0$ & always &  \\
\hline
\end{tabular}
\tabcolsep 0pt \caption{\label{Tab1} The critical points of the autonomous system.}
\vspace*{5pt}
\end{table} 

\begin{figure}[htbp]
\centering
\includegraphics[scale=0.8]{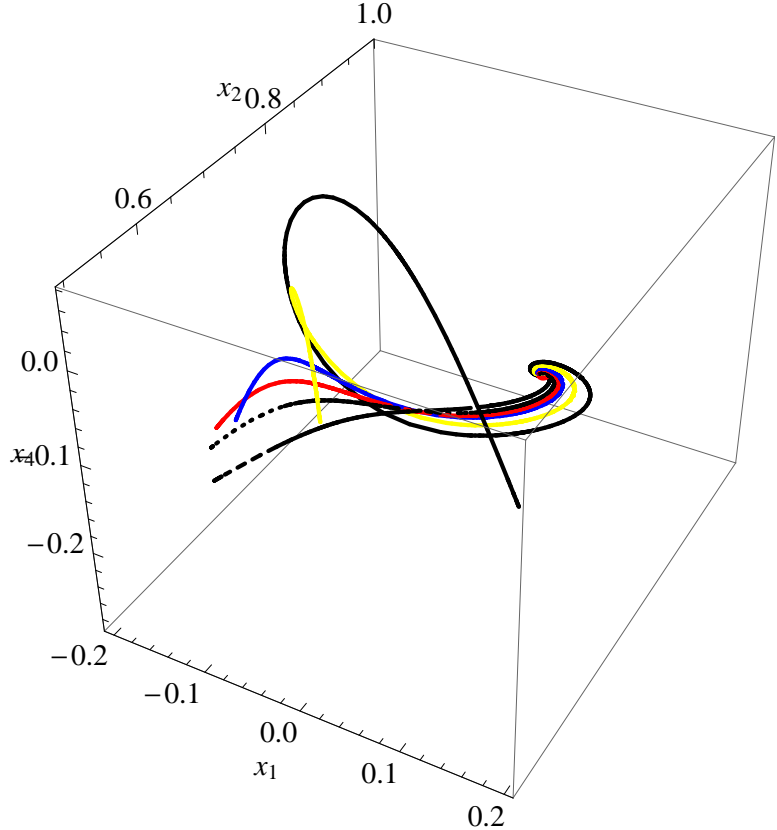}
\caption{\label{Fig2}The phase diagram of $(x_1, x_2, x_4)$ with different initial conditions. }
\end{figure}

Now we study the stabilities of the critical  points for the autonomous system (Eq.~(24)-Eq.~(27)). Considering linear perturbations and linearizing the autonomous system (Eq.~(24)-Eq.~(27)), we  obtain four differential equations. The eigenvalues of the coefficient matrix of these  differential equations determine the stability of the critical points. If the real parts of all eigenvalues are negative, the corresponding critical point is stable, otherwise, it is unstable.  After some tedious calculations, we get four eigenvalues for a given critical point and show them in Tab.~(\ref{Tab2}). Apparently, only Point (C.P.a) is stable, and both Points (C.P.b) and (C.P.c) are unstable. Thus,  there is only one attractor for the system, which is a dark energy dominated de Sitter solution.
In Fig.~(\ref{Fig2}), we plot a phase diagram of $(x_1, x_2, x_4)$ with different initial conditions, which shows that all curves converge into a point, which is just the stable point (C.P.a). Therefore, our universe will enter finally  a de Sitter expansion phase.

\begin{table}[!h]\centering
\begin{tabular}{|c|c|c|c|c|c|c|c|}
\hline
Label &~ The eigenvalues  ($\lambda _1,\lambda _2,\lambda _3,\lambda _4$)~&~Stability~\\
\hline
C.P.a & $-3,\;0,\;-\frac{3}{2}+\frac{3}{2} \sqrt{1-\frac{8}{3} \sqrt{\omega  (-\lambda ^2+4 \omega )}},\;-\frac{3}{2}-\frac{3}{2} \sqrt{1-\frac{8}{3} \sqrt{\omega (-\lambda ^2+4 \omega )}} $& Stable \\
\hline
C.P.b  &$-3,\;0,\;-\frac{3}{2}+\frac{3}{2} \sqrt{1+\frac{8}{3} \sqrt{\omega  (-\lambda ^2+4 \omega )}},\;-\frac{3}{2}-\frac{3}{2} \sqrt{1+\frac{8}{3} \sqrt{\omega  (-\lambda ^2+4 \omega )}}$  & Unstable  \\
\hline
C.P.c    &$3,\;\frac{3}{2},\; \frac{1}{4} (-3-\sqrt{9+48 \omega }),\; \frac{1}{4} (-3+\sqrt{9+48 \omega })$ & Unstable \\
\hline
\end{tabular}
\tabcolsep 0pt \caption{\label{Tab2} The eigenvalues of critical points.}
\vspace*{5pt}
\end{table}

\section{Conclusion}
The extended quintessence dark energy is obtained by coupling a normal scalar field to the Ricci scalar defined in the metric formalism and it has an advantage of realizing the crossing of phantom divide line compared with the quintessence which can not.
In this paper, we propose a new extended quintessence dark energy by also introducing a non-minimal coupling between the quintessence scalar field and gravity, but  with the Ricci scalar given from the Palatini formalism rather than the metric one. Although in the minimal coupling case the field equation in  the Palatini formalism is the same as that in the metric formalism, when a non-minimal coupling is introduced, the results are wholly different.  We find that the equation of state of the new extended quintessence can cross the phantom divide line and oscillates around the $-1$ line. This oscillatory behavior is different from that of the extended quintessence and teleparallel dark energy where only one crossing of $-1$ line occurs. In addition, we find that the universe will enter a dark energy dominated de Sitter phase in the future.  This result is further confirmed by the dynamical analysis, which shows that there is only one attractor which corresponds to a dark energy dominated de Sitter solution.

\begin{acknowledgments}

This work was supported by
the National Natural Science Foundation of China under Grants Nos.
10935013, 11175093, 11222545 and 11075083, Zhejiang Provincial Natural Science
Foundation of China under Grants Nos. Z6100077 and R6110518, the
FANEDD under Grant No. 200922, the National Basic Research Program
of China under Grant No. 2010CB832803, the NCET under Grant No.
09-0144,  and K.C. Wong Magna Fund in Ningbo University.
\end{acknowledgments}

\end{document}